\documentstyle[aps,prb,multicol]{revtex}

\begin{document}
\draft

\title{\bf Dynamical mean-field theory of the small polaron}

\vskip 2cm
\author{S. Ciuchi$^{\dagger}$,
F. de Pasquale$^{\P}$,
S. Fratini$^{\P *}$,
D. Feinberg$^{\S}$}

\address{{$^\dagger$} Dipartimento di Fisica, Universit\`a de
L'Aquila, via Vetoio, I-67100 Coppito-L'Aquila, Italy}
\address{$^{\P}$ Dipartimento di Fisica,
Universit\`a di Roma "La Sapienza",
piazzale Aldo Moro 5, I-00185 Roma, Italy}
\address{$^{\S}$ Laboratoire d'Etudes des Propri\'et\'es Electroniques
des Solides$^{**}$, Centre National de la Recherche Scientifique,  BP 166 38042
Grenoble Cedex 9, Grenoble, France}
\date{\today}
\maketitle

\begin{abstract}
A dynamical mean-field theory of the small polaron problem is presented,
which becomes exact in the limit of infinite dimensions. The ground state
properties and the one-electron spectral function are obtained for
a single electron interacting with Einstein phonons by a mapping of the
lattice problem onto a polaronic impurity model. The one-electron
propagator of the impurity model is calculated through a continued fraction
expansion (CFE), both at zero and finite temperature, for any
electron-phonon coupling and phonon energy. In contrast to the ground state
properties such as the effective polaron mass, which show a continuous behaviour
as the coupling is increased, spectral properties exhibit a sharp qualitative
change at low enough phonon frequency: beyond a critical coupling, one energy
gap and then more and more open in the
density of states at low energy, while the high energy part of the spectrum
is broad and can
be qualitatively explained by a strong coupling adiabatic approximation.
As a consequence
narrow and coherent low-energy subbands coexist with an incoherent featureless
structure at high energy. The subbands denote the formation of quasiparticle
polaron states. Also, divergencies of the self-energy may occur in the gaps. At
finite temperature such effect triggers an important damping and broadening of
the polaron subbands.  On the other hand, in the large phonon
frequency regime such a separation of energy scales does not exist and the
spectrum has always a multipeaked structure.

\end{abstract}



\section{Introduction: the single polaron problem}

The polaron problem is an old but not fully solved problem
of solid state physics. The {\it small polaron} theory which will be considered
here assumes a short-range electron-phonon interaction and explicitly includes
the lattice periodicity \cite{Tiablikov,Holstein}. It is therefore aimed to
study systems where screening is effective. This addresses for instance the
situation of a metal consisting of different bands, one of them being narrow
enough to allow for a strong coupling to phonons. In fact, if the crystal can
be considered as made of strongly deformable molecular-like units with
narrow-band electrons hopping from one to another, then the conditions for
a strong polaron effect can be realized \cite{Holstein}. In realistic
structures, for example transition metal oxides or organic metals, such units
exist which provide local (oscillation) phonon modes and are indeed strongly
coupled to well-defined electronic orbitals. Recently, much interest has
been revived into polaron theory, due to important classes of materials,
including the high-temperature superconductors\cite{HTSC} and the ''colossal''
magnetoresistance manganites\cite{Millis,Tyson,Yamada}.
In fact, in the insulating parent phase of superconducting
cuprates, polarons have been unambiguously detected by optical
measurements \cite{Kim,Taliani,Calvani}, and some evidence of strong
electron-phonon coupling effects has been given recently in the metallic
phase\cite{Calvani,Bianconi}. In the manganites, on the other hand, strong
static or dynamic Jahn-Teller distorsions appear at the
metal-insulator transitions \cite{Millis,Tyson,Yamada}.

The problem of a single polaron becomes relevant for low carrier density, but
also by itself, as a paradigm to study the effect of strong coupling
electron-phonon interactions. In the intermediate and strong coupling
regimes, the small polaron problem is already a non trivial many-body
problem\cite{Landau-Pekar}: the difficulty consists in describing the dressing
of the electron by a coherent multiphonon cloud, moving coherently with it such
as to form a quasiparticle. Perturbative techniques, starting either from
the free electron limit or from the atomic limit (strong-coupling
expansions)\cite{Holstein,Lang-Firsov,Alexandrov-Ranninger,Gogolin,Alexandrov}
fail in describing the dressing effect in the intermediate regimes. On the other
hand, nonperturbative solutions based for instance on variational ansatz
\cite{LeeLowPines,Emin,lavorone} are expected to give reliable results only for
the ground state properties like the effective polaron mass. But no
satisfactory description of the full spectral properties has been obtained so
far.

Holstein's Molecular Crystal Model\cite{Holstein} involves tight-binding
electrons coupled to dispersionless optical phonons. As a function of phonon
frequency and electron-phonon coupling, it displays a variety of interesting
regimes. The strong coupling regime leads to the formation of small polarons,
with a dramatic increase of the effective mass for low phonon
frequencies\cite{Holstein,Lang-Firsov}. On the other hand, for exactly zero
phonon frequency, an adiabatic solution can be obtained with self-trapping of
the polarons appearing only above some critical coupling value, in dimensions
greater than one \cite{Emin,Kabanov-Mashtakov,Lagendijk1}.

In this context, the recent discovery of a nonperturbative theory for
interacting quantum problems, based on the limit of infinite lattice
coordination
(or dimensionality) \cite{Georges et al.,Jarrell,Van Dongen,Kotliar}, opened a
new way of attacking strong electron-phonon problems. A few recent papers have
addressed the problem of superconductive and charge-density wave instabilities
of the metallic state, close to half filling, either in weak
coupling\cite{Ciuchi}, using self-consistent techniques
\cite{Amalfi,Freericks-conserving}, or from the local impurity method
\cite{FJS}, also including a local electron repulsion (the Holstein-Hubbard
problem)\cite{FJ}. Also, a solution at finite density in the adiabatic limit
(zero phonon frequency) was obtained \cite{Millis-Shraiman}. On the other hand,
exact results for the spectral function of a single polaron at zero temperature
were recently reported by the
authors\cite{Ciu95}. The aim of the present work is to provide a complete
description of the small polaron crossover based upon the knowledge of spectral
quantities in the whole range of parameters and also at non zero temperature.
The most striking features are found in an intermediate coupling regime where no
known approximation scheme works.

Similarly to the Hubbard model, the infinite
dimensional limit allows to map the lattice problem onto a self-consistent local
impurity model, here called "polaron impurity"\cite{Ciu95,FJS}. It
consists of a single-site electron-phonon problem, embedded into a quantum
effective medium characterized by an effective ''free'' propagator which has to
be  self-consistently determined. This mapping preserves all the complexities of
the quantum dynamics of the problem, namely the interplay between electron
and lattice fluctuations {\it at the local level}.
The crucial point here
is that for a single electron the impurity model can be {\it analytically}
solved by a recursion formula for any non-interacting impurity  propagator,
leading to a {\it continued-fraction expansion} (CFE) solution for the fully
dressed propagator. This unique feature allows to obtain at the end an
exact solution for the lattice problem in the limit of infinite dimensions.
This solution provides directly, with modest
computational efforts, the ground state as well as the spectral properties
in the thermodynamic limit and at any temperature.
Therefore it is somehow complementary to
numerical works performed in finite dimensions, such as Monte Carlo
simulations\cite{Lagendijk1,Lagendijk2} or exact diagonalization of
finite clusters\cite{Marsiglio,Grilli}. Indeed, the former are limited
to finite temperatures, and the latter have to deal with finite size
effects.

The main
result of the present self-consistent impurity approach is that, in the
crossover regime, low and high energy scales can be accurately described,
like in the Mott-Hubbard transition problem\cite{Kotliar}. Polaron
states in the low energy range appear as coherent strongly renormalized
quasiparticle states, while at higher energies the electron is incoherently
scattered by a quasiclassical random distortion. These features are clearly
displayed in the low phonon frequency regime, where the spectral density
displays low energy peaks coexisting with a broad and incoherent high energy
continuum.  This provides a novel and physically transparent
representation of the polaron crossover. As the coupling strength is
increased, it proceeds through successive opening of gaps in
the spectral density, separating polaron subbands. At intermediate
couplings, and decreasing the phonon frequency, one finds that the dressing
of electron states by a multiphonon coherent cloud drastically reduces the
effective electronic energy scales and leads to an adiabaticity
"catastrophe" in the low energy spectrum. In terms of a perturbative
expansion, this can be clearly ascribed to high order vertex dressing. It
is important to emphasize that even in an intermediate regime of couplings
a well defined polaron quasiparticle excitation is present at low energy.
Also, an important feature is the appearance of a discrete set of frequencies
where the self-energy diverges   within the low-energy gaps of
the spectrum. Those points are
sensitive to disorder\cite{Hotta}. In particular, at finite
temperatures, they enhance the damping and thermal broadening
of polaron states in their vicinity, leading to a loss of coherence of
the main polaron subband.

The paper is organized as follows . In Section II, we
introduce the Holstein molecular crystal model and we discuss the
main limiting cases in a finite dimensional lattice.
In Section
III, we introduce the impurity analogy and the exact CFE solution of the
impurity model, at zero and at finite temperature.
In the same section the limiting results of the CFE are presented and compared
to the finite dimensional case.
Section IV presents the general results of the CFE solution of the impurity
problem. Section V is devoted to conclusions and to the discussion of possible
extensions of this method to finite density, in relation with a Coherent
Potential Approximation (CPA) formulation of the CFE.

\section{The Holstein model}

In this section we shall summarize the main results concerning the Holstein
model in finite dimensions, in
the limit in which analytical calculations can be performed. We especially focus
on the role of dimensionnality, which will allow to discuss later how our theory
(exact in infinite dimensions) can be compared with results in finite
dimensional
lattices.

The Holstein model consists of tight-binding conduction electrons
interacting with local dispersionless phonon modes. The Hamiltonian is
\begin{equation}
\label{Holstein-model}
H = - \sum_{<ij>,\sigma} t_{i,j}
(c_{i,\sigma}^\dagger  c_{j,\sigma} + h.c.) - g \sum_{i,\sigma}
c_{i,\sigma}^\dagger c_{i,\sigma} (a_i + a_i^\dagger ) + \omega_0 \sum_i
a_i^\dagger a_i
\end{equation}
where $c_{i,\sigma}^\dagger (c_{i,\sigma})$ creates
(destroys) an electron with spin $\sigma$ at site $i$, and $a_i^\dagger
(a_i)$ creates (destroys) a phonon at site $i$. The hopping matrix elements
$t_{i,j}$
connect nearest neighboring sites of a lattice in $d$ dimensions and
we assume they give rise to a band of half-bandwidth
$t$.
This model possesses
two independent control parameters\cite{lavorone}. The first one is the
bare coupling constant $\lambda = g^2/\omega_0 t=|\epsilon_p|/t$, where
$\epsilon_p=-g^2/\omega_0$ is the polaron energy obtained in the atomic
limit ($t=0$). The second one is the adiabatic parameter $\gamma =
\omega_0/t$. A third parameter can be conveniently introduced as a
combination of the above ones, as $\alpha = g/\omega_0$, with $\alpha^2 =
\lambda/\gamma$. While $\lambda$ and $\gamma$ are commonly used as
parameters in the perturbative analysis, the parameter $\alpha$ which
measures the strength of the lattice deformation involved in the polaron
effect will show to be crucial in the strong coupling regime. Let us
stress that these parameters are defined from the bare energy scales
$\omega_0, g, t$ in the Hamiltonian, contrarily to usual definitions in the
theory of electron-phonon interaction in metals where in particular the
dressed phonon frequency is used.
It is worth defining the following
regimes and limits which are relevant to the Holstein model

i) weak (strong) coupling $\lambda<1$ ($>1$)

ii) small (large) phonon frequency $\gamma<1$ ($>1$)

iii) multiphonon regime $\alpha^2>1$

iv) adiabatic limit $\omega_0=0$, finite $\lambda$.

Fig.  1 shows the corresponding
regions in the ($\lambda, \gamma$) plane.

Throughout this paper we shall concentrate on the problem of one
electron in interaction with phonons, i.e. a system which density is
zero in the  thermodynamic limit. To analyze the perturbative behaviour of
the model we shall first discuss the
simplifications due to this limit.
The discussion is restricted to  the zero temperature limit but
can be easily generalized to finite temperatures (see section
II B).
For a single electron the
Green's function in the site representation can be defined as
\begin{equation}
\label{Green-def}
G_{i,j}(t)=-i \langle 0 | T c_j(t) c_i^\dagger (0) | 0 \rangle
\end{equation}
where $|0\rangle$ is the vacuum for phonons and electrons
and the unessential spin indices are omitted. One observes
that there is only one possible ordering ($t>0$) of the T-product, so that
the function is purely retarded\cite{Mahan}. Then the standard
perturbation theory is introduced in the site representation by defining
the electron self-energy $\Sigma_{i,j}(\omega)$\cite{Metzner-Vollhardt}
through the Dyson equation
\begin{equation}
\label{Dyson-lattice}
G_{i,j}=[{\cal G}_0]_{i,j} + \sum_{k,l} [{\cal G}_0]_{i,k} \Sigma_{k,l}
G_{l,j}
\end{equation}
where ${\cal G}_0$ is the free electron propagator.

At zero density, the following simplifications hold:
\begin{itemize}
\item[-]{A general self-energy diagram consists of a single electron
line first emitting and then absorbing phonons.}
\item[-]{The emission (absorption) of a phonon consists in
subtracting (adding) a quantum of phonon frequency $\omega_0$ to
the energy of the propagating electron line.}
\end{itemize}

The first statement comes from the absence of
density fluctuations in the zero density limit  (no bubble diagrams, no phonon
renormalization).
Moreover in the zero temperature limit all the phonons must be created from
vacuum before being absorbed.
To illustrate the second statement, we first notice that in a
generic self-energy diagram involving $N$ phonon lines, it is always possible
to choose an integration contour which avoids all the poles and cuts
of the electron Green's function,
since the retarded electron
propagator is analytic in the upper half plane.
Then the only contribution comes
from the poles associated to  the phonon lines.

\subsection{Weak coupling and adiabatic limit}

The perturbation expansion of the self-energy to
second order in $g$ gives a local ({\bf k}-independent) self-energy
\begin{equation}\label{2nd-order}
\Sigma_2(\omega)=g^2 {\cal G}_0 (\omega-\omega_0)
\end{equation}
where ${\cal G}_0$ is the local free propagator obtained by the knowledge
of the free particle DOS $N(\epsilon)$
as ${\cal G}_0=\int d\epsilon N(\epsilon) (\omega-\epsilon)^{-1}$.
Notice that
dimensionality enters only through the free DOS \cite{Engelsberg}.
The electron effective mass, in the case of a local self-energy,
is easily calculated via
\begin{equation}
\label{eff-mass-def}
\frac{m^*}{m} = 1-{\frac{d Re \Sigma(\omega)}{d \omega}}|_{E_0}
\end{equation}
where $E_0$ is the ground state energy.

Let us first consider the low phonon frequency regime.
In this case for electron states lying at the bottom of
a band, dimensionality effects enter through the band shape near the band
bottom and control both the behaviour of the effective mass and
the spectral properties.
We assume that near the band bottom
($\epsilon=-t$) $N(\epsilon)\sim(1+\epsilon/t)^{d/2-1}/t$, then one has from
eqs. (\ref{2nd-order},\ref{eff-mass-def}) for $0<d<4$\cite{adiabatic}
\begin{equation}
\label{eff-mass-ad}
\frac{m^*}{m}=1+\lambda k_d \gamma^{d/2-1}
\end{equation}
where $k_d$ is a numerical constant. \cite{kd}
It is not surprising that we
do not recover the expected "Migdal" result $1+\lambda$.
In fact this last result is obtained assuming an infinite flat band which could
be the case of a metal which Fermi energy lies far from  singularities (like Van
Hove ones) in the DOS.  From eq. \ref{eff-mass-ad} we can define
an effective coupling $\tilde{\lambda}=\lambda N(E_0+\omega_0)t$
as in the case of Van-Hove singularities\cite{Baffetto-Pietronero} in order to
write $m^*/m=1+\tilde{\lambda}$.  The effective coupling
strength $\tilde{\lambda}$ tends to zero for vanishing phonon
frequency in dimensions $d>2$ (keeping $\lambda$ constant), while it
goes to infinity for $d<2$. Surprisingly, a {\it perturbative} analysis
provides non trivial information about the adiabatic limit:
for $d>2$ we expect a free electron behaviour while for $d<2$ the
perturbation expansion {\it around a delocalized solution}
fails in the adiabatic limit for any finite $\lambda$.
This is consistent
with the non-perturbative findings of ref.\cite{Kabanov-Mashtakov},
where it is shown that
renormalization effects are absent up to a {\it finite} value
of $\lambda=\lambda_c$ for $d\geq 2$, while in $d=1$ the behaviour is
polaronic for any finite value of $\lambda$\cite{adiabatic2}.

In the opposite case of large phonon frequency,  calculating the
self-energy in eq. (\ref{2nd-order}) for large $\omega_0$ and taking advantage
of the asymptotical behaviour of the free propagator at high energy,
it is easy to get
\begin{equation}
\label{eff-mass-antiad}
\frac{m^*}{m}=1+\alpha^2
\end{equation}
It appears as a first order expansion in power of $\alpha^2$. As we shall
see this expansion is actually resummed, in the large phonon frequency limit,
by the use of the Lang-Firsov\cite{Lang-Firsov} transformation.

It is worth comparing the spectral properties obtained at finite bandwidth
and zero density with those derived
up to second order in $g$ in the classical work of  Engelsberg and
Schrieffer\cite{Engelsberg} in the case of an infinite bandwidth.
Since the self-energy is local one can define the spectral function as a
function of energy $\epsilon$ and frequency
$\omega$
\cite{Engelsberg}
\begin{equation}
\label{spectral-function}
A(\epsilon,\omega) = -\frac{1}{\pi} Im \frac{1}{\omega-\epsilon-\Sigma(\omega)}
\end{equation}
where $\epsilon$ runs over the non-interacting band energies $\epsilon_{\bf k}$
of a translationally invariant lattice.
The electron spectral density ${\cal N}(\omega)=-(1/\pi) Im G(\omega)$ is
derived from eq. (\ref{spectral-function}) by integrating over the energy
distribution.

In agreement with ref.
\cite{Engelsberg}, one easily obtains a quasiparticle excitation
spectrum  (with $Im\Sigma=0$) at energies $E_0\le\omega\le E_0+\omega_0$
and an incoherent broad spectrum at larger energies.

In the low phonon frequency case the spectral density of the low-energy
quasiparticle states can be determined by the low-energy properties
of the free DOS. A band of coherent excitations could separate from an
incoherent states, depending on
the dimensionality.
The equation which determines the band edges is
\begin{equation}
\label{ground-state-eq}
\omega-Re \Sigma(\omega) = \pm t
\end{equation}
The minus sign determines the band bottom {\it including the ground state
energy}
while the plus sign determines the band top. Using the self-energy of eq.
(\ref{2nd-order}) it is easy to see that near $E_0+\omega_0$ ,  $Re\Sigma$
diverges in $d=1$ and  $d=2$ as $\omega^{-1/2}$ and $\log(\omega)$ respectively,
while it  is well behaved for $d>2$. Uneven with ref.\cite{Engelsberg} the
finite  bandwidth effects taken into account
by eq. (\ref{ground-state-eq}) generate a gap in $d=1,2$.
More precisely the amplitude of the gap (in units of the bandwidth)
between coherent and incoherent
states scales with $(\lambda\gamma)^4$ in $d=1$ and with
$\exp(-1/\lambda\gamma)$ in $d=2$.
In $d=3$ the real part of the self-energy does not diverge at
$E_0+\omega_0$ so  that a sufficiently large coupling is necessary
to fulfill eq. (\ref{ground-state-eq}).
In this case we expect the appearance of a gap only for $\lambda$ greater than
a certain value which depends on the phonon frequency and explicitly on the
whole band shape.

The discussion of this subsection suggests that, in dimensions larger than 2,
perturbative expansions fail beyond some critical coupling above which gaps open
up in the one-electron density of states. This property will be revealed in
detail by the self-consistent local impurity theory analyzed in sections III,IV.

\subsection{Atomic and large phonon frequency limits}

The {\it atomic} limit is defined as the zero hopping case ($t = 0$).
It can be understood also as an infinite coupling limit
$\lambda\rightarrow\infty$.
One considers a single electron on a single site lattice (atom) whose
Hamiltonian is given by eq. (\ref{Holstein-model}) with $t=0$.  In the case
of zero bandwidth the Hamiltonian of eq. (\ref{Holstein-model})  can be
diagonalized by the unitary Lang-Firsov (LF)  transformation\cite{Lang-Firsov}
\begin{equation}
U = exp[\alpha c^{\dagger} c (a-a^{\dagger})]
\end{equation}
The effect of this transformation is to shift the phonon operators by a
quantity $\alpha$ such
as the electron-phonon interaction is eliminated. It introduces a new
fermion, the  polaron, which carries this phonon field shift
\begin{equation}
c\rightarrow X c
\end{equation}
where $X=\exp{\alpha(a-a^{\dagger})}$. Once the transformation is performed, the
Hamiltonian becomes diagonal and the ground state energy is the polaronic
energy $\epsilon_p = -g^2/\omega_0$, the excited polaron states having an energy
$\epsilon_p+n\omega_0$.

Due to the presence of an electron at a given site, the lattice is
deformed. The magnitude of this effect is measured by the local part of the
static electron-displacement correlation function defined as
\begin{equation}
\label{def-X}
C_0 = \langle n_i (a_i +a_i^\dagger ) \rangle
\end{equation}
In the atomic limit one gets $C_0 = 2 \alpha$ which means that the atomic ground
state is that of a localized polaron i.e. an electron surrounded by a "cloud"
represented by a coherent (Glauber) phonon state, with an average number
$<a^\dagger a>=\alpha^2$ of phonons.  The electron propagator can also be
calculated after the LF transformation\cite{Mahan}  \begin{equation}
G(\omega) = \sum_{n=0}^\infty
\frac{\alpha^{2n}e^{-\alpha^2}}{n!}\frac{1}{\omega-n\omega_0-\epsilon_p}
\label{atomicGreen}
\end{equation}
The resulting spectral density appears as a Poissonian distribution of
delta peaks separated by the phonon frequency $\omega_0$.
By exploiting the Lehmann representation of the Green's
function one can see that such a distribution is due to the
projection of a localized zero phonon state onto the (n-phonon)
polaron eigenstates of
the Hamiltonian. This can be usefully understood from a gedanken X-ray or
optical absorption experiment, where the wavefunction of the
localized electronic final state (with undistorted lattice) is expanded onto the
(lattice relaxed) polaron eigenstates, which builds the electron spectral
function. From eq. (\ref{atomicGreen}) we see that the Green's function has
a spectral weight close to the ground state energy which is
exponentially small in the interaction strength, while the spectral weight is
maximum   for excitations involving approximately $n \sim \alpha^2$ phonons.

Let us now consider the action of the hopping. An
approximation valid for large phonon frequencies is derived from the
LF transformation, applied to the Hamiltonian with a non zero
hopping term. The hopping term, modified by the transformation,
represents the hopping of the {\it polaron}
\begin{equation}
t_{i,j} c_{i,\sigma}^\dagger c_{j,\sigma}\rightarrow t_{i,j}
X^{\dagger}_i X_j
c_{i,\sigma}^\dagger c_{j,\sigma}
\end{equation}
The Holstein approximation\cite{Holstein} consists in averaging the
polaron kinetic energy on the free phonon variables, thus obtaining at zero
temperature an effective hopping amplitude
\begin{equation}
t_{i,j} \langle 0|X^{\dagger}_i X_j|0\rangle = t_{i,j} e^{-\alpha^2}
\end{equation}
for $i,j$ nearest neighbors.
This approximation amounts to neglect phonon emission and absorption during the
hopping process. It is believed to give correct results  when $\omega_0$ is the
largest energy scale\cite{Baffetto}.

In the same spirit, following Alexandrov and Ranninger, one can go further and
use the same approximation to calculate the {\it electron} propagator, also for
finite electron density\cite{Alexandrov-Ranninger}, which gives
\begin{equation}
\label{HolsteinGreen}
A(\epsilon_k,\omega) = -\frac{1}{\pi} Im \left[
\frac{e^{-\alpha^2}}{\omega-\epsilon_k^*-\epsilon_p}+\sum_{n=1}^{\infty}
\frac{1}{N} \sum_q
\frac{\alpha^{2n}e^{-\alpha^2}}{n!}
\frac{1}{\omega-\epsilon_q^*-n\omega_0-\epsilon_p} \right]
\end{equation}
where $\epsilon_k^*=\epsilon_k \exp(-\alpha^2)$ runs over the renormalized
bandwidth obtained by replacing the free hopping parameter $t$ by
$t^*=t\exp(-\alpha^2)$. This solution shows
a coherent low energy quasiparticle band
describing a polaron of effective mass
\begin{equation}
\label{eff-mass-Holstein}
\frac{m^*}{m} = e^{\alpha^2}
\end{equation}
located around $\epsilon_p$, together with an
incoherent structure at higher energies.

Let us give a physical interpretation of this result by showing that at
least for the low energy states it corresponds to substituting the exact
self-energy with the atomic one. This is valid in the case $\omega_0>>t$ where a
generic scattering process will lead electrons through intermediate states
out of the band. In this scattering process the system can be thought as a
flat band "atomic" system in interaction with high energy phonons.
For frequencies near the polaron ground state energy $\epsilon_p$, the atomic
self-energy reads
\begin{equation}
\label{sigma-at}
\Sigma(\omega) = \omega (1-e^{\alpha^2})+e^{\alpha^2}\epsilon_p
\end{equation}
Using the definition eq. (\ref{spectral-function}) we
get the spectral function
\begin{equation}
A(\epsilon,\omega)=-\frac{1}{\pi}Im
\frac{e^{-\alpha^2}}{\omega-\epsilon^*-\epsilon_p}
\label{Aholstein}
\end{equation}
where $\epsilon^*=\epsilon \exp(-\alpha^2)$ describes the renormalized band.
Thus one recovers the low energy part of eq. (\ref{HolsteinGreen})
from an approximation to
the self-energy which is  justified in the large phonon
frequency regime and {\it near the quasi-particle polaronic peak}.
However, for $\alpha^2>>1$, i.e. when multiphonon effects are important, the
validity of the Holstein approximation is questionable
even in the case of large {\it but finite} phonon frequency.
Results from small cluster exact diagonalization\cite{Ranninger-Mello}
show that the adiabatic ratio $\omega_0/t$ must increase as
$\alpha^2$ to  ensure the validity of the Holstein
approximations (see fig. 1).

Finally, let us give the result for the electron
propagator in the atomic limit, at finite temperature. The Green's function
is then defined generally by averaging on
phonons only (the problem is that of a "cold" electron in a thermalized
phonon bath). In the atomic limit, it is obtained in the same way as at
$T=0$, yielding the pole representation\cite{Mahan}
\begin{equation} G(\omega) =
\sum_{n=-\infty}^{+\infty} e^{-(2N+1)\alpha^2} I_n\{ 2 \alpha^2
[N(N+1)]^{1/2} \} e^{n\omega_0/2T} \frac{1}{\omega-n\omega_0-\epsilon_p}
\label{atomicGreenT}
\end{equation}
where $N=exp(-\omega_0/T)$ is the phonon thermal
weight and $I_n\{ z\}$ are the Bessel functions of complex argument.

Comparing this expression with eq. (\ref{atomicGreen}), we remark that at
$T\neq 0$ the corresponding spectral function displays peaks at frequencies
below the polaron ground state energy, with a spectral weight
which is exponentially small at low temperatures. This apparent paradox of
having electron states at {\it lower} energy than the ground state
can be explained if one interprets these states as polaron states formed
after absorbing $n$ thermal phonons from the thermal bath, with a
probability $exp(-\beta n \omega_0/2)$. This reduces the cost in lattice
energy required to form the polaron. Since the polaron energy results from
a balance between this (positive) cost and the (negative) electron-lattice
coupling energy, it is possible to create states lying below the zero
temperature ground state level. The price to be paid is that these states are
incoherent, due to the incoherent (thermal) phonon distribution. Also,
the chemical potential goes to minus infinity, allowing the
fermion occupation number to be zero at any energy for one particle at
finite $T$.

\section{The impurity analogy and the exact solution for
a single electron}

The dynamical mean field theory is developed as the exact solution of an
infinite dimensional\cite{Muller-Hartmann} or infinite connectivity lattice.
It has been shown\cite{Metzner-Vollhardt,Muller-Hartmann}
that to have a finite free electron kinetic
energy the hopping matrix elements must be scaled with the square root of
the lattice dimensionality or lattice coordination\cite{dinf-phonons}.
A second point to deal
with is the proper choice of the
infinite coordination lattice in order to get a finite value of the ground state
energy. In fact, a problem arises for example in the case of an hypercubic
lattice,
which has a gaussian DOS with an infinite tail towards low energy
\cite{Muller-Hartmann}. For large but finite dimensions the scaling of the
hopping matrix elements implies that the ground state energy of one electron
is proportional to $\sqrt{d}$. Therefore, the formation of a small
polaron requires an electron-phonon coupling energy of the same order of
magnitude and a coupling constant $\lambda$ which diverges with
$\sqrt{d}$. Indeed, just like the formation of a bound state from an
external potential, polaron formation by self-trapping requires an infinite
coupling strength in an ordinarily connected lattice in infinite dimensions
(see ref. \cite{Ciu95}). To overcome this difficulty we consider a Bethe
lattice of infinite coordination $d$. The hopping matrix elements in eq.
(\ref{Holstein-model}) have been scaled as $t_{i,j}=t/2\sqrt{d}$ with $t$
being the  half-bandwidth of the lattice. In the Bethe lattice (see
ref. \cite{Economou}, par. 5.3.4) only
self-retracing paths are allowed. A restriction of the possible paths
to go from one site to another allows this lattice to mimic a finite
dimensional one also in the limit of infinite coordination, giving rise
to a finite bandwidth semi-elliptical free DOS
\begin{equation}
\label{Bethe-DOS} N(\epsilon)=\frac{2}{\pi t^2}\sqrt{t^2-\epsilon^2}
\end{equation}
which correctly simulates the low energy features of a three dimensional
lattice. More generally, in interacting fermion problems, with this particular
choice, localization phenomena can be
found even in the infinite dimension limit, for instance the Mott-Hubbard
transition is correctly obtained at a finite coupling for the half-filled
Hubbard
model\cite{Kotliar}.

Let us now come to the essential simplification occurring in the
limit of infinite dimensions, namely the fact that
the electron self-energy is {\it local} in space.
The usual
argument which holds for local electron interactions
\cite{Muller-Hartmann,Jarrell,Vollhardt}
can be worked out also in the context of the Holstein model. One can carry on
the standard argument but taking into account also the phonon self-energies and
electron-phonon vertices instead of the electron-electron four leg vertices
\cite{Muller-Hartmann}. This gives the scaling of the real space
propagator with the intersite ''Manhattan'' distance $R=|i-j|$ as $G(R)\propto
1/\sqrt d^{R}$. Using this scaling and the skeleton expansion of the vertex
function  one can prove that the self-energy is not only local but that
it  {\it depends only on local phonon and electron propagators}
\cite{Muller-Hartmann,Kotliar}.
In practice in a generic self-energy diagram all the internal lines are local
propagators.

Using the locality of the self-energy, the lattice propagator in the $k$-space
is then given by $G_{\bf k}(\omega) = [\omega - \epsilon_{\bf k} -
\Sigma(\omega)]^{-1}$ where $\epsilon_{\bf k}$ is the tight-binding
electronic dispersion. Writing $G_{ii} = 1/N\sum_{\bf k} G_{\bf k}$ and
introducing the free DOS as $N(\epsilon)=1/N\sum_{\bf k}
\delta (\epsilon-\epsilon_{\bf k})$
one has (dropping the site index for $i=j$)
\begin{equation}
\label{lattice-self-cons}
G(\omega)=\int d{\epsilon}
\frac{N(\epsilon)}{\omega-\Sigma(\omega) - \epsilon}
\end{equation}
Notice that in infinite dimensions the properties of the lattice enter only
through the free electron DOS.

Having a local self-energy, one can
demonstrate following ref.\cite{Jarrell} the existence of an impurity model
equivalent to the lattice problem. This can be readily seen by writing the
real-space Dyson equation for the {\it local} propagator $G_{ii}$ (eq.
(\ref{Dyson-lattice})) in two steps. The first one involves self-energy
contributions on sites $j$, with $j \neq i$. Once these contributions are
resummed, one is led with a modified local propagator, noted $G_{0}$. The latter
can be used in the full Dyson equation for $G_{ii}$, reintroducing the missing
self-energy contributions which involve {\it the same site i}. It leads to the
{\it local} Dyson equation for $G_{ii} = G$ (assuming
translational symmetry)
\begin{equation}
\label{imp-self-cons}
G(\omega) = [G_0^{-1}(\omega) - \Sigma(\omega)]^{-1}
\end{equation}
The problem is therefore that of an impurity embedded into a
medium. All the electron-phonon scattering processes occurring on
sites other than the impurity site are contained in the effective
"free" impurity propagator $G_0$, while local processes at the
impurity site are taken into account by the self-energy $\Sigma$ in eq.
(\ref{imp-self-cons}).

This impurity problem can be made more physical by parametrizing it as a
"polaron" Anderson impurity model involving a localized "$d$" level coupled to a
local phonon, and hybridized with a fictitious conduction electron band "$c$" of
dispersion $E_{\bf k}$
\begin{equation}
\label{Himp}
H_{imp}=
\sum_{k} E_{\bf k} c_{\bf k}^\dagger c_{k} -
\sum_{k} V_{\bf k} (c_{\bf k}^\dagger d +
d^\dagger c_{k})
+ \omega_0 a^\dagger a
- g  d^\dagger d (a + a^\dagger )
\end{equation}
with new impurity parameters $V_{\bf k}$ and $E_{\bf k}$ being related to the
propagator $G_0$ by
\begin{equation}
G_0^{-1}(\omega) = \omega - \int_{-\infty}^{+\infty} d E
\frac{\Delta(E)}{\omega- E}
\end{equation}
and
\begin{equation}
\Delta(E) = \frac{1}{N}\sum_{\bf k}
V_{\bf k}^2 \delta (E - E_{\bf k})
\end{equation}
The original {\it lattice} Green's function it is that of the $d$ level.
Therefore, solving the problem defined by the impurity Hamiltonian of eqs.
(\ref{Himp}) for a given $G_0$ and applying the self-consistency conditions
eqs. (\ref{lattice-self-cons},\ref{imp-self-cons}) one has the so called
Local Impurity Self-consistent  Approximation (LISA) which is the exact
solution of a $d\rightarrow\infty$ problem. Interestingly enough, the
above impurity Hamiltonian has been used in the past to model core level
relaxation in the X-ray problem\cite{Cini}.  However, in the context of the LISA
approach, its significance becomes much more general. Just as the repulsive
Anderson impurity model for the Hubbard model, it plays the role of a "paradigm"
impurity model for the physics contained in the Holstein Hamiltonian.
Though the advantages of using an impurity parametrization of the
$d\rightarrow\infty$
problem  have been extensively reported in ref.\cite{Kotliar} (we also
refer to the original references), we must stress, as a general fact, that
such a parametrization is not unique in the LISA context.

\subsection{The zero temperature formalism}

In the case of one single electron the Green's function at zero
temperature in terms of the
impurity operators, is
\begin{equation}
\label{Gret}
G(t)=-i\theta(t) \langle0| d(t) d^\dagger (0) |0\rangle
\end{equation}
This function
describes the  propagation amplitude of the impurity electron created
from the vacuum at
time zero and destroyed at time $t$. Fourier transforming eq.
(\ref{Gret}) leads to the resolvent
\begin{equation} \label{Gfreq}
G(\omega)=\langle0|d\frac{1}{\omega+i\delta-H}d^\dagger|0\rangle
\end{equation}
which has the correct prescription $\delta>0$ for convergence of the time
integrals. The vacuum energy is defined here to be zero.

Solving the lattice problem requires to find the solution of the impurity
problem for any given $G_0$.
Let us separate the impurity Hamiltonian of eq. (\ref{Himp}) into  $H_0$ and
$H_I$, where $H_0$ alone leads to the effective free propagator $G_0$ and $H_I$
is the local interaction term, then  an operator identity for the resolvent
holds
\begin{equation} \label{Risolv}
\frac{1}{z-H}=\frac{1}{z-H_0}+
                   \frac{1}{z-H_0} H_I \frac{1}{z-H}
\end{equation}
The diagonal matrix element of this operator on the impurity
zero phonon state
$d^\dagger |0\rangle$ is the Green's function of eq.
(\ref{Gfreq}). To proceed further one needs to introduce the
generalized matrix elements\cite{Cini}
\begin{equation}
\label{matrix-elements}
G_{n,m}=\langle0|\frac{a^n}{\sqrt{n!}}d \frac{1}{\omega+i\delta-H}d^\dagger
\frac{(a^\dagger)^m}{\sqrt{m!}}|0\rangle
\end{equation}
so that the element $G_{0,0}$ will be the solution of the $T=0$ problem.

In the case of $H_I$ given by (\ref{Himp}), to express the matrix element of
the right term in eq. (\ref{Risolv}) one takes advantage of the linearity
of the interaction term in the electron density operator $n=d^\dagger d$.
Namely,
introducing a set of zero electron $p$-phonon states
$|0,p\rangle=(a^\dagger)^p/\sqrt{p!}|0\rangle$
one can write
\begin{equation}
H_I=\sum_p d^\dagger |0,p\rangle\langle 0,p|d (a+a^\dagger)
\end{equation}
leading to the recursion formula for the $G_{n,m}$'s
\begin{equation} \label{Eqric}
G_{n,m}=G_{0n} \delta_{n,m} -g\sum_p G_{0n} X_{n,p} G_{p,m}
\end{equation}
where  $G_{0n}=G_0(\omega-n\omega_0)$ is the diagonal element of the
free resolvent and $X_{n,p}$ are the phonon displacement matrix
elements
\begin{equation}
X_{n,p}= \sqrt{p+1} \delta_{n,p+1}+ \sqrt{p} \delta_{n,p-1}.
\end{equation}
Eq. (\ref{Eqric}) is solved in matrix notation
\begin{equation}
\label{G-matrix}
{\bf G}^{-1}={\bf G_0}^{-1}+g{\bf X}
\end{equation}
One immediately recognizes that, due to the particular form of ${\bf X}$,
${\bf G}^{-1}$ is a tridiagonal matrix, so
that the solution of the problem is reduced to the inversion of a matrix
in arbitrary dimensions.
Following the lines given in ref.\cite{Viswanath-Muller}
(see alternatively ref. \cite{Cini}) one can
express the diagonal element of the ${\bf G}$ matrix in terms of the
diagonal and non-diagonal elements of ${\bf G}^{-1}$.
The local propagator (the $0,0$ element of ${\bf G}$) is obtained in
terms of a Continued Fraction Expansion (CFE), as a functional
of the ''bare'' propagator $G_0$:
\begin{equation}
\label{CF}
G(\omega)={1 \over\displaystyle G_0^{-1}(\omega)-
{\strut g^2 \over\displaystyle G_0^{-1}(\omega-\omega_0)-
{\strut 2g^2 \over\displaystyle G_0^{-1}(\omega-2\omega_0)-
{\strut 3g^2 \over\displaystyle G_0^{-1}(\omega-3\omega_0)-...}}}}
\end{equation}
Due to the impurity analogy,
{\it this is also
the local propagator of the original lattice problem}, provided
that eqs. (\ref{lattice-self-cons},\ref{imp-self-cons})
are fulfilled. As a special case, one notices that in the atomic limit,
setting $G_0(\omega)=\omega^{-1}$, eq.(\ref{CF}) is nothing but an
alternative representation of the atomic propagator, eq.(\ref{atomicGreen}).
In the general case, the self-energy is immediately recognized as a
functional of
$G_0$, from the self-consistency condition (\ref{imp-self-cons})
\begin{equation}\label{CF-sigma} \Sigma(\omega)=\displaystyle {\strut g^2
\over\displaystyle G_0^{-1}(\omega-\omega_0)- {\strut 2g^2 \over\displaystyle
G_0^{-1}(\omega-2\omega_0)- {\strut 3g^2 \over\displaystyle
G_0^{-1}(\omega-3\omega_0)-...}}} \end{equation}
This allows to solve the impurity problem in the dynamical mean
field theory. Once the self-energy is obtained for a given $G_0$, the local
lattice propagator is calculated from eq. (\ref{lattice-self-cons}) and
using eq. (\ref{imp-self-cons}) a new $G_0$ is obtained. After few
numerical iterations a fixed point $G_0^*$ is reached, and the lattice
local propagator $G^{-1}=G_0^{*-1}-\Sigma[G_0^*]$ is determined. We
emphasize that eq. (\ref{CF-sigma}) only involves a {\it discrete} set of
values of $G_0$ and this turns out to be a drastic simplification of the
calculation. Moreover, in the Bethe lattice the impurity
and the local propagator are simply related:
\begin{equation}
\label{G0-bethe}
G_0^{-1}(\omega)=\omega-\frac{t^2}{4}G(\omega)
\end{equation}
which, replacing eqs. (\ref{lattice-self-cons},\ref{imp-self-cons}),
simplifies the calculation.

Let us now show diagrammatically how formula (\ref{CF-sigma}) exactly sums
up all
of the self-energy contributions. It is indeed possible to relate term by term
the expansion of eq. (\ref{CF-sigma}) in powers of $g$ considering $G_0$ as a
parameter in the skeleton expansion of the self-energy. The relation is
obtained through the following steps:

i) first obtain a truncation to a given formal order $g^k$ of the skeleton
expansion of the self-energy using the rules previously introduced.

ii) express the internal {\it fully interacting} propagator of the
skeleton expansion in terms of the "bare" impurity propagator
$G_0$ using the self-consistency
conditions (\ref{lattice-self-cons},\ref{imp-self-cons})
and expand the result to $g^k$.
The result is equal to the expansion of the continued fraction\ to the {\it
formal}
order $g^k$.

It is now instructive to understand the meaning of a finite
truncation of the CFE.
The self-energy given in eq. (\ref{CF-sigma}) can be defined recursively
\begin{equation}\label{se-rec}
\Sigma^{(p-1)}(\omega)=\frac{pg^2}{G^{-1}_0(\omega-p\omega_0)-\Sigma^{(p)}
(\omega)}
\end{equation}
where $p$ is the stage index of the CFE  for $\Sigma$. An $N$-stage
truncation of the CFE is defined by neglecting $\Sigma^{(N+1)}$ in eq.
(\ref{se-rec}). In the resulting diagrammatic expansion {\it only phonon
states $|n\rangle$ with $n\leq N$ appear as intermediate states}. This
operates a selection of diagrams which is different from that based on the
perturbative expansion, which by contrast is related to the number of
interaction vertices. Indeed, at each step of the truncation an infinite set of
diagrams is resummed, {\it including vertex corrections} (see fig. 2).  As one
can easily see by writing the expansion for the atomic limit, the parameter
$\alpha^2$ measures the importance of multiphonon effects and the number of
phonons $N$ needed for an accurate description of all the scattering processes
should be much larger than $\alpha^2$. It is interesting to note that a
self-consistent non-crossing approximation such as a Migdal scheme, always
fail, since it can be put in a CFE context by changing all coefficients of $g^n$
in eq. (\ref{CF-sigma}) to $1$.

\subsection{Generalization to a thermalized lattice}

The above formalism can be easily generalized to
non zero temperature. The trace performed over free phonon states in
gives
\begin{equation}
\label{Gfreq-T}
G(\omega) = (1-e^{-\beta \omega_0}) \sum_n e^{-\beta
n \omega_0} G_{n,n}(\omega)
\end{equation}
where $G_{n,n}$ are the diagonal elements of the correlation matrix
defined in eq. (\ref{matrix-elements}) and calculated by means of
the Dyson eqs. (\ref{G-matrix}).

The calculation of the diagonal elements $G_{n,n}$
follows the lines given in refs. \cite{Viswanath-Muller,Cini}.
The inverse of each
$G_{n,n}(\omega)$ is now the sum of an {\it infinite}
continued fraction which is similar to the result at $T=0$ plus a
{\it finite} fraction which formally takes into account the
absorption processes at negative frequencies. This reads
\begin{equation}
\label{Gnn}
G_{n,n}(\omega)={1 \over\displaystyle G_0^{-1}(\omega)-
A-B}
\end{equation}
where
\begin{equation}
A=
{\strut ng^2 \over\displaystyle G_0^{-1}(\omega+\omega_0)-
{\strut (n-1)g^2 \over\displaystyle G_0^{-1}(\omega+2\omega_0)-
{\strut (n-2)g^2 \over\displaystyle \ddots -
{\strut g^2 \over\displaystyle G_0^{-1}(\omega+n\omega_0)}}}}
\end{equation}
and
\begin{equation}
B=
{\strut (n+1)g^2 \over\displaystyle G_0^{-1}(\omega-\omega_0)-
{\strut (n+2)g^2 \over\displaystyle G_0^{-1}(\omega-2\omega_0)-
{\strut (n+3)g^2 \over\displaystyle
G_0^{-1}(\omega-3\omega_0)- \cdots}}}
\end{equation}
The solution of the problem now follows the same lines of the zero
temperature case.

The relation between the CFE expansion and the perturbation theory
can be exploited using the simplification which holds in the zero
density limit at non zero temperatures.
The rules for constructing a self-energy diagram at non zero temperature
are easily obtained as generalization of those obtained in section II B.

\begin{itemize}
\item[-]{A general self-energy diagram consists of a single electron
line emitting and absorbing phonons.}
\item[-]{The emission (absorption) of a phonon consists in
subtracting (adding) a quantum of phonon frequency $\omega_0$ to
the energy of the propagating electron line. Associate to each process a
factor $1+f_B(\omega_0)$ when subtracting $\omega_0$ and a factor
$f_B(\omega_0)$ when adding $\omega_0$ to the electron
line.}
\end{itemize}

The introduction of the temperature energy scale $T/\omega_0$ rules the
truncation of the series of CFE eq. (\ref{Gfreq-T}). For any finite $n$ we
also consider, in the practical evaluation of the spectral function, a
finite truncation of $B$ of size $N$. We therefore have a maximum number of
phonons $N+n$ in an intermediate virtual state of which $N$ are emitted and
$n$ are absorbed from the thermal bath. Therefore the criterium of
truncation valid at $T=0$ ($N>>\alpha^2$) has to be supplemented by the
condition $n>>T/\omega_0$.

\subsection{Limiting cases in the LISA approach}

We discuss here some limiting cases based upon the LISA approach and show that
the main properties of the finite dimensional polaron problem are captured
by the
infinite dimensional limit. We also analyze the adiabatic limit which, when the
polaron becomes localized, involves the breaking of translational symmetry and
consequently cannot be achieved using the CFE, which assumes this symmetry.

By expanding to second order the LISA self-energy given by eq. (\ref{CF-sigma})
and substituting the free propagator for the self-consistent $G_0$  we
obtain the perturbative relation of eq. (\ref{2nd-order}).
Then from eq.(\ref{eff-mass-def}) and using
the semi-elliptical (Bethe lattice) DOS of eq.(\ref{Bethe-DOS})
one gets the effective mass
\begin{equation}
\frac{m^*}{m}=1-2\lambda \gamma
[1-\frac{1+\gamma}{\sqrt{(1+\gamma)^2-1}}]
\end{equation}

In the low phonon frequency regime, this becomes
\begin{equation}
\label{eff-mass-ad-dinf}
\frac{m^*}{m}=1+\sqrt{2\gamma} \; \lambda
\end{equation}
which shows the same behaviour as in a regular three dimensional case (see eq.
\ref{eff-mass-ad}). In the large phonon frequency regime we obtain the same
result as eq. (\ref{eff-mass-antiad}).

The fourth order term in the self-energy expansion is
\begin{equation}
\label{4nd-order}
\Sigma_4(\omega)=2 g^4 {\cal G}_0^2 (\omega-\omega_0) {\cal G}_0
(\omega-2\omega_0)
\end{equation}
>From the previously stated rules, $\Sigma_4$ is the sum of two
contributions, e.g. a fourth order non-crossing diagram (fig. 2
b))plus a vertex correction (fig. 2 c)).  In the zero density limit
and in infinite dimensions, {\it the two contributions are exactly
equal}. As a consequence, any non-crossing approximation such as
the self-consistent approximation of Engelsberg and
Schrieffer\cite{Engelsberg} is not valid here. In fact, this kind of
approximation can be justified by the the conventional Migdal's
argument\cite{Migdal} restated for Einstein phonons\cite{Engelsberg}.
This argument requires the condition $\lambda \omega_0/E_F \ll 1$
where $E_F$ is the Fermi energy \cite{Marsiglio}, which is trivially
invalidated in the zero density limit where $E_F=0$.

Concerning the spectral properties most of the perturbative
consideration of the three dimensional case (section II) are valid.
Due to the fact the the real part of $\Sigma$ to second order does not
diverge, a finite gap arises in the spectral density for
sufficiently large values of the phonon frequency
($\gamma>2/(1+2\lambda)$).

In the atomic limit, all the self-consistent $G_0^{-1}(\omega-n\omega_0)$ in
eq. (\ref{CF}) must be substituted by their atomic value
$\omega-n\omega_0$.  The resulting continued fraction can be also
independently obtained by solving directly the atomic model
through the resolvent
technique described in the previous section.
The advantage of this formulation, compared to the LF result
(eq.(\ref{atomicGreen})) is to yield  immediately the self-energy as a
functional of the free atomic propagator. The exact result at finite hopping
could then be understood as the result of a Coherent Potential Approximation
(CPA) procedure. By CPA we mean the self-consistent approximation which amounts
to substituting in the CFE form of the atomic self-energy a self-consistent
Green's function for the atomic one \cite{Vollhardt}. It is worth noting that
although the "pole" and CFE expressions of the atomic propagator are equivalent,
they give different results when extended to finite hopping through the CPA.
For example if one extracts the self-energy for the CPA procedure
from eq.  (\ref{atomicGreen}), one recovers an expression  which agrees
with the exact results only for large phonon frequencies.

The Holstein approximation described in section II B can be recovered at low
energy by the CFE expansion of the local propagator when $\omega_0$ is the
largest energy scale in the self-consistent propagator $G_0$.
In this case, all the
$G_0^{-1}(\omega-n\omega_0)$ with $n\ge0$ appearing in eq. (\ref{CF})
can be replaced by their atomic value giving an atomic self-energy
which yields the exponential renormalization of the effective mass
 as shown in section III.

Another instructive formula can be derived from the CFE in the limit
$\omega_0\rightarrow 0$.
In this case all the $G_0(\omega-n\omega_0)$
in eq. (\ref{CF}) can be replaced by $G_0(\omega)$ and one
recognizes the continued fraction expansion of the
complex error function\cite{Abramowitz} which can be
expressed in terms of
an integral
\begin{equation}
\label{adiabatic self}
     G(\omega)= \int \frac{dx}{\sqrt{2\pi}} e^{-x^2/2}
     \frac{1}{G_0(\omega)-gx}
\end{equation}
The physical interpretation of this formula is that the electron moves
within a field of displacements with a gaussian distribution. Such distribution
can be understood as the "classical" limit of the quantum probability
distribution of local lattice displacements when the phonon frequency goes to
zero while keeping the elastic energy finite. Let us remark that as far as the
hopping self-energy term in eq. (\ref{adiabatic self}) is neglected one recovers
a gaussian DOS as predicted for  $\omega_0\rightarrow0$  by the atomic limit.

The adiabatic limit is reached when both $\omega_0$ and $g$ go to zero, keeping
$\lambda$ fixed. In this case the CFE yields the free electron propagator. The
problem is that one needs to consider the possibility
of translational symmetry breaking.
We have developed an independent
scheme presented in Appendix A which
allows an exact
solution at zero phonon frequency {\it keeping $\lambda$ finite}.
The ground state energy is determined by
minimizing the total energy with respect to the lattice displacement,
described by a
classical variable. According to the shape of the total energy curve as a
function of the lattice displacement (see fig. 3) we find three different
regimes

i) $\lambda<\lambda^\prime_c$. The only stable
minimum corresponds to an undistorted lattice (delocalized solution,
strictly free electron).

ii) $\lambda^\prime_c<\lambda<\lambda_c$. The delocalized solution is still a
stable minimum, but a relative minimum appears in the potential at non zero
lattice  deformation, corresponding to a metastable localized solution
(small polaron).

iii) $\lambda>\lambda_c$. The stable minimum corresponds to a localized
solution.

where $\lambda^\prime_c=0.649..$ and $\lambda_c=0.843..$.

Therefore in the $d\rightarrow\infty$ Bethe lattice a first-order
localization transition occurs at $\lambda_c$ from a delocalized free electron
to a localized polaron.
Moreover, for finite values of the coupling, the
localized polaron extends over several lattice shells of neighbors around a
given localization site, just like in finite
dimensions\cite{Kabanov-Mashtakov}.
All these features are quite similar to those found in regular 2-d and 3-d
cubic lattices.

To summarize the discussion of this section, examination of the various
limiting regimes in the special case of an infinite dimensional Bethe
lattice shows the consistency of this limit with a three dimensional
situation. This is true as well in the adiabatic regime as in the
perturbative and large phonon frequency regimes (shaded areas in fig. 1).
The dynamical mean field solution, exact in
infinite dimensions and presented in the following section, allows us to
complete the phase diagram and can be thought as a controlled
interpolation scheme valid at least qualitatively also in finite dimension
$d>2$.

\section{Results from the dynamical mean-field theory}

Let us now turn to the self-consistent solution for the
infinite-dimensional lattice with a semi-elliptical density of states, by
solving eqs. (\ref {G0-bethe}, \ref {CF}). We first discuss the ground
state properties, as deduced from the behaviour of $\Sigma(\omega)$ close
to the ground state energy.

\subsection{Ground state properties}

The knowledge of the self-energy and of the Green's function
allows to access to the ground-state
properties: the ground state energy, which is evaluated by solving equation
 (\ref{ground-state-eq});
the electron-lattice {\it local} correlation function defined in eq.
(\ref{def-X}) which
can be evaluated using the Hellmann-Feynman theorem
\cite{Hellmann-Feynman} from the first derivative of the ground state
energy with respect  to $g$;
the electron kinetic energy i.e. the average over the ground state of the
hopping term of the Hamiltonian eq. (\ref{Holstein-model})
which, using again the
Hellmann-Feynmann theorem, is calculated as a derivative of $E_0$
with respect to $t$;
the average phonon number in the ground state
which can be  obtained as a derivative
of the ground state energy with respect
to $\omega_0$.

The ground state properties are summarized in figures
 4, 5, 6, 7, 8. They illustrate the above relevant quantities
as functions of the coupling constant
$\lambda$ for three different values of $\alpha^2=1,2,5$. The polaron
crossover is seen as a continuous change from weakly dressed to quasi localized
electrons. The crossover almost disappears  for $\alpha^2=1$ while it
becomes sharper for large $\alpha^2$,  approaching the first-order localization
transition observed in the adiabatic limit  $\omega_0=0$ (see Appendix A). The
existence of a smooth crossover rather than an abrupt transition (for
finite $\omega_0$) corroborates the general proof previously derived by
Gerlach and L\"owen\cite{Gerlach}. However, the ground state properties
are quantitatively calculated here for the first time for any value
of the parameters.

Let us discuss these results in more details. Concerning the ground state
energy  depicted in fig. 4 one sees that it is
bounded from above by the adiabatic result ($\omega_0=0,\alpha^2\rightarrow
\infty$) and from below by the large phonon frequency result
($\omega_0=\infty,\alpha^2=0,E_0=-\lambda$). As the coupling increases, the
crossover occurs for $\lambda$ of the order of $\lambda_c$, where
$\lambda_c$ is the critical coupling strength obtained in the  adiabatic
limit. The behaviour of the effective mass is shown in fig.
 5. For large $\lambda$ it increases with $\alpha^2$, but
remains smaller than Holstein's prediction of eq.
(\ref{eff-mass-Holstein}), which is attained asymptotically only for very
large couplings\cite{alpha2-fix}.

Notice also that for $\alpha^2=1$ we do not observe any change in the
curvature of the effective mass, showing that no appreciable crossover
occurs but rather a smooth increase of the effective mass from 1 towards
$exp(1)\sim 2.7$. On the other hand, one notices that for
$\lambda<\lambda_c$, the effective mass diminishes as $\alpha^2$ increases,
in agreement with the adiabatic ($\alpha^2 \rightarrow \infty$) prediction
of having unrenormalized electrons for small couplings. Therefore, the mass
renormalization as a function of $\alpha^2$ behaves in opposite ways for
$\lambda<\lambda_c$ and $\lambda>\lambda_c$. For infinite $\alpha^2$
($\omega_0=0$), $m^*/m$ jumps from $1$ to $\infty$.

The spectral properties,
discussed in the next paragraph will clarify this singular behaviour which
indeed reflects the breakdown of the perturbation theory for
$\lambda>\lambda_c$.

The ground state kinetic energy is shown in fig.
 6. Again a crossover is found as a change of curvature
only for $\alpha^2>1$ and becomes sharper as the  adiabatic limit is
approached. Figure 7 displays the electron-phonon
correlation function (i.e. the local deformation of the lattice). To make a
comparison with the well defined adiabatic limit, it is convenient to scale
this quantity by the strong coupling value $2\alpha$. A sharp crossover
towards large electron-lattice local correlations is found for large
$\alpha^2$. Similarly, the number of phonons in the ground state, shown in
fig. 8 attains the value $\alpha^2$ only
asymptotically for very large coupling. As a general property, one must
stress that the Holstein values for all the above quantities are obtained
assuming a {\it local} lattice deformation. The gradual behaviour we find
towards these values is due to the finite extension of the polaron, i.e. of
the electron wavefunction and lattice deformation over several shells of
lattice neighbours. This is also true in the $d\rightarrow\infty$ limit, and in
particular causes the kinetic energy to be non zero, even in the adiabatic limit
for $\lambda>\lambda_c$ (see appendix A).

Recently, a numerical study\cite{Grilli} by diagonalization on small
clusters has led to the conclusion that the polaron crossover occurs when both
conditions $\alpha^2>1$ and $\lambda>1$ are fulfilled. According to this
interpretation, to have a polaron one requires that $\lambda>1$ for
$\gamma<1$ or $\alpha^2>1$ for $\gamma>1$. CFE results are in qualitative
agreement with this statement since we observe no appreciable crossover for
any value of $\lambda$ provided $\alpha^2<1$, and in the opposite case the
crossover is found at around $\lambda\sim 1$ and becomes sharper as
$\alpha^2$ is increased.
A better understanding of this behaviour can be gained
by plotting the effective mass in the whole parameter space
($\gamma$, $\lambda$) fig. 9 a). The isolines
corresponding  to large effective mass  define the polaron region. We see
that for large $\lambda$ and $\gamma$ the effective mass depends only on
$\alpha^2$ as predicted by the
strong coupling theory. As $\gamma$ decreases, the crossover gets sharper
 until it becomes a real first order localization transition
for $\gamma=0$
at the adiabatic critical value $\lambda_c$. Finally it must be remarked
that all these results are in qualitative agreement with Monte-Carlo
simulations\cite{Lagendijk1,Lagendijk2},
which show that {\it at finite phonon frequency} the ground state properties
in the polaron crossover are not much dependent on the dimensionality.

\subsection{Spectral properties at $T=0$}

As in other strong coupling problems such as the Hubbard model, the
standard mean-field or variational techniques (based here on the Lang-Firsov
approximation followed by phonon averaging) do not allow to go beyond the low
energy properties. Instead, the dynamical mean-field theory provides a way to
explore the whole electron spectrum.

The spectral properties are directly extracted from the knowledge of
the  local propagator, and reflect the structure of the excited states.
Formulas  (\ref{spectral-function},\ref{ground-state-eq}), valid in
the case of a local self-energy, obviously apply in the case of the dynamical
mean-field theory.

We focus here on the behaviour of the
spectral density and of the self-energy. In particular, a nonzero
$Im\Sigma$ reveals an incoherent scattering due to emission and absorption
of phonons. More precisely, given a band in ${\cal N}(\omega)$ of reduced width
$t^*$ we can determine if an excitation at a given frequency has a coherent
or an incoherent character. To do this we expand the spectral function
around the peak located at $\omega^*-Re\Sigma(\omega^*)=\epsilon$. This
relation defines a pole $\omega^*(\epsilon)$ which,
in the case of a translationally invariant lattice,
gives the band dispersion. Then assuming around the
pole $\omega^*$ an effective mass inversely proportional to the effective
bandwidth and using eq. (\ref{eff-mass-def}) with
$E_0\rightarrow\omega^*$ we get approximately
$\omega-\epsilon-Re\Sigma(\omega)\simeq (t/t*)(\omega-\omega^*)$ and
\begin{equation}
A(\epsilon,\omega)\simeq -\frac{1}{\pi} \frac{t^*}{t}
\frac{\Gamma}{(\omega-\omega^*)^2+\Gamma^2}
\end{equation}
where $\Gamma=-(t^*/t)Im\Sigma(\omega^*)$. Then the excitation is coherent
if its lifetime ($1/\Gamma$) is much greater than the
characteristic time of the (renormalized) hopping processes $1/t^*$ i.e.
$\Gamma \ll t^*$. This gives the coherence condition
$Im\Sigma(\omega^* ) \ll t$ (and not $t*$). If the coherence condition holds,
the quasiparticle pole $\omega^*(\epsilon)$
is well defined since the spectral function has a sharply defined
peak at $\omega=\omega^*(\epsilon)$ with a width much less
than the renormalized bandwidth.

The scenario for polaron formation can be analyzed from figs.
 10, 11, 12 where the spectral
density and the imaginary part of the self-energy are shown in the large,
intermediate and low phonon frequency regimes respectively, for
increasing values of $\lambda$. Generally speaking, in all regions of
parameters,  the condition to form a (quasi)particle of large
effective mass such as a small polaron is that {\it a narrow
coherent band emerges at low  frequency}. We call it the zero-phonon
{\it polaronic band}. It is worth noticing that when a polaron with
a large effective mass is formed one always observes
several bands in the spectra, and the one at lowest energy is {\it
perfectly coherent} ($Im \Sigma = 0$), since this band lies entirely below
the minimum energy $E_0+\omega_0$ for inelastic scattering\cite{Engelsberg}.
Though this
feature is common to the spectra in all parameters' regions, the
way in which a polaronic behaviour shows in the spectral
properties is very different according to the value of
the adiabatic ratio $\gamma$.

Let us first discuss the large phonon frequency regime. In this case (see
fig.  10) the formation of the polaron from the
point of view of the spectral properties is a smooth crossover.
For $\gamma>2$, the spectra always display a multipeaked structure
(see also fig.  9 b)) in which the $n=0$ polaron band and
the edges of the
first excited bands are perfectly coherent. These structures are
subbands corresponding to polaron states with $n$ phonons excited,
and can be easily understood by switching on the hopping term from the atomic
limit. For very large $\gamma$, all the
subbands tend to have the same width, in agreement with the
predictions from Holstein's approximation\cite{Alexandrov-Ranninger}, see eq.
(\ref{HolsteinGreen}).  For intermediate values
of $\gamma$, increasing $\lambda$
causes the shrinking of the width of the low energy bands and decreases their
spectral weight, while higher frequency bands become more important. The
crossover occurs for $\lambda \sim  \gamma$ ($\alpha^2 \sim 1$)
(cf.  12-c in which $\alpha^2=2$).  One notices that the
envelope of the peaks in the imaginary part of the self-energy tends to
reproduce the envelope of the bands in the spectral density, shifted by
$\omega_0$, as can be deduced from the CFE.  Thus, for $\alpha^2>1$, both
the weight and
the damping of the subbands tend to increase with their index and then
decrease following a roughly  Poissonian envelope, with a maximum at
$\omega\sim \alpha^2\omega_0+E_p \simeq 0$.

Let us now turn to the low phonon frequency regime, in which the formation
of polaronic bands is qualitatively different and exhibits novel features.
>From figure 12 we see that a polaronic band emerges from an
{\it incoherent} band around $\lambda\simeq 1$. Increasing the value of the
coupling, more and more bands emerge, having very small bandwidth. Notice
that the low energy structures
are not resolved on the scale of fig. 12 c), but they can in fact be accurately
calculated by the CFE and are shown in detail in fig. 13. As we shall
see below, the relative distance between polaron subbands is expected to be
less than $\omega_0$, due to lattice displacements effects. The inverse lifetime
$Im\Sigma$, on the opposite, reproduces the pattern of the polaron subbands at
energies shifted by $\omega_0$. Therefore, if the bandwidth renormalization
is strong
enough, each $n$-th order excited band splits further into a doublet of
bands separated by a gap, where the main one is coherent and the secondary
one is incoherent.  The coherent subbands turn out to have equivalent
heights and
can be  interpreted as coherent quantum tunneling out of a distorted
lattice site: they
correspond to {\it coherent polaron} bands with $n=0, 1, ...$ excited phonons.

A qualitative understanding of the low and high energy
excitations at small $\gamma$ can be deduced from a comparison with
the adiabatic limit results (appendix A). Fig. 14
shows the spectrum of fig.  12 c), in correspondence with
the adiabatic potential relative to the same value of the coupling
$\lambda$. It is clear from figs. 14
that a low energy scale can be defined as the region where the
spectrum consists of separated subbands. The separation of energy scales can be
understood by considering the effect of finite phonon
frequency i.e. by considering
quantum corrections to the classical lattice approximation.
This yields a series of low energy  bands which are roughly
centered around the position of the quantized  levels of the
ground-state adiabatic potential $V^{(b)}_{ad}(X)$ given in eqs.
(\ref{E-loc1},\ref{E-loc2}) and fig.  14. As the energy is increased
from $E_0$,
hese levels become more and more hybridized
with the excited adiabatic continuum associated to the
undistorted lattice
starting at $V^{(a)}_{ad}(0)=-t$.
In the real spectrum, one actually observes that the
very narrow bands merge into a broad structure just around this
energy level, thus defining the  amplitude of the low energy
region as $|E_0|-t$.
One can also evaluate from the data shown in figs. 12 c), 13, 14 a
small negative
deviation of about 2\% of the first band spacings from $\omega_0$,
and in general the first four narrow bands in the figure  are not
exactly equally spaced. This can be explained in terms of the
adiabatic potential picture by noticing that the curvature of the
total adiabatic energy near the distorted minima is smaller than
near the undistorted position. In fact, a prediction based upon
linearization of the adiabatic potential yields a 1.65\% deviation
from $\omega_0$ for the  distance of the first excited level from the
ground state of the adiabatic  potential.

The nature of the high energy part of the spectrum is explained using a
strong coupling adiabatic approximation, namely the
$\omega\rightarrow0$ limit of the CFE\cite{strong-adiabatic} (see
eq. \ref{adiabatic self}). This
approximation describes the broad structure as an envelope
of resonances separated by a vanishingly small $\omega_0$.
The resulting spectrum is shown in fig  14 (upper panel) and fits
very well the high energy part of the spectral density. This fit
deteriorates at intermediate energies where spiky structures appear, due to
the effect of a finite $\omega_0$. The self-energy obtained from (\ref{adiabatic
self}) indicates that all the states  at intermediate and high energies have
an incoherent character.

In the intermediate frequency regime, the structure of the low energy bands
when these are well separated is curious and follows some rules that can be
deduced directly from  the CFE expansion (see fig. 11 d)):
the lowest energy band is always  coherent ($Im \Sigma=0$) and its
shape resembles the original semi-elliptical unrenormalized DOS even if
some asymmetry is observed towards its upper edge. This is not obtained by
the usual (Holstein) strong coupling approximation, and could be depicted
by a much larger effective mass at the top band-edge. On the other hand, the
higher order bands acquire a complexity which can be labelled by the number
of "substructures" that can be recognized in the band shape and increases
with increasing energy.
Moreover, by comparison of figures  12 and 11 we see
that the transition with increasing $\lambda$ from a single band structure
to a multi-peaked structure at low frequencies is much more
abrupt for
small than intermediate $\gamma$, and in the former case it turns out to
occur around the adiabatic critical value of the coupling $\lambda_c$.

Let us now focus in more detail on the mechanism for gap opening in the low
and intermediate phonon frequency regimes ($\gamma<2$). The number of
well-separated subbands is shown in fig. 9 b)
as a function of
the parameters $\lambda$ and $\gamma$. One notices that there is a large
region of the parameters' space in which the spectral density consists of
only two structures: a single coherent polaronic band and a high energy
band which is mostly incoherent.
In this region, approximation
(\ref{HolsteinGreen}) which, apart from the low-energy coherent peak,
displays an incoherent spectrum made of separated subbands, is not correct.
The extent of this region reduces as the
adiabatic ratio is reduced and for very low phonon frequency the system
undergoes a rapid crossover to a multi-peaked structure at around
$\lambda_c$.
The emergence of a coherent band, separated by a finite energy gap from the
continuum can be seen as $\lambda$ increases on fig.  15.
As a critical value of $\lambda$ is approached, a pseudogap appears
together with a large damping of the states in the same energy range. At
higher $\lambda$, the gap is formed and keeps increasing until a second gap
is formed at higher energy, and so on. By the way, let us
underline that the total spectral weight carried by the polaron bands, when
they are well separated, is in general different from that obtained in the
atomic limit. For instance, the $n=0$ band spectral weight is always
larger than $e^{-\alpha^2}$.

An effect related to the formation of subbands,
which arises in the intermediate coupling regime, is the
 divergency of the self-energy at one and
subsequently more and more frequencies located in the gaps
as the coupling strength is increased.
In the presence of disorder, external excitations or at finite temperature,
 a non zero spectral density can appear
around these energies. We thus expect such excited
states within the gaps to have a huge damping
and to be localized, while the ground state can keep its
delocalized character.
For that motivation the authors\cite{Hotta} who found such behaviour in
the context of the Holstein-Hubbard model named it
{\it dynamical localization}.
We must stress that the relevance of such a phenomenon concerning the mobility
properties such as the ac conductivity cannot be tested using the CFE formalism.
However, we draw attention to the
fact that this is a very general phenomenon, which occurs here for a single
electron as a polaronic feature i.e. is related to the multi-peaked
structure of the Green's function.

The occurrence of such self-energy divergencies requires that the
states at the edges of two consecutive bands have an infinite
lifetime ($Im \Sigma=0$) at zero temperature.
In this case, the self-energy fulfills the
eq. (\ref{ground-state-eq}) at the extrema of the energy
gap, so that the real part of $\omega-\Sigma$ changes sign
within the gap  without crossing the values $\pm t$. As a consequence
a divergency in $Re \Sigma$ occurs at a point $\omega_L$
located in the gap.
In the Bethe lattice we have the condition
\begin{equation}
\label{dyn-loc}
\omega_L-\frac{t^2}{4}G(\omega_L-\omega_0) - \Sigma^{(2)}(\omega_L) = 0
\end{equation}
where $\Sigma^{(2)}(\omega_L)$ is the second stage expansion of the self-energy
in the  CFE of eq. (\ref{se-rec}).
Equation (\ref{dyn-loc}) may have more than one solution, i.e. many
such points are expected for large values of
$\lambda$ (see fig.  13).
In the atomic limit ($\lambda\rightarrow\infty$) there is an infinite
number of them, located in the gaps between the  peaks
of the atomic spectral density.

In the ($\lambda$,$\gamma$) plane, at least one self-energy divergency
exists on the right of the  dashed line in fig.  9
b).
We note that a self-energy divergency is
formed only in a preexisting energy gap.
Accordingly, when increasing the coupling strenght a gap appears first,
separating a coherent polaronic peak from an incoherent excited band, and then a
self-energy divergency occurs in the gap when the states at the bottom of the
excited band become coherent.

This phenomenon should affect the spectral properties in
the presence of disorder\cite{Hotta} or at finite temperatures (see
below).
In fact, whenever there is a mechanism which is able to give rise to
excited states in the proximity of those special points (this can
be due to the lorentzian tails induced by disorder) the singularity in the
imaginary part of the self-energy becomes a finite broadening lorentzian,
which implies a loss of coherence of the neighboring states. The consequences on
the electronic spectra at finite temperatures will be analyzed in the following
section.

\subsection{Spectral properties  at finite temperatures}

In this section we present the results obtained at finite temperature using
the CFE formulation of eq. (\ref{Gfreq-T}) and following.
The analysis focuses on the intermediate regime where no "classical" schemes of
approximation are  available.

In Holstein's original treatment of polaron motion\cite{Holstein,Mahan},
a distinction is made between transition amplitudes which are {\it
diagonal} in the phonon number, contributing to the polaron coherent motion
around the ground state energy,
and those which are {\it non-diagonal}, giving rise to hopping-like motion.
While the former decrease in the presence of thermal disorder, the latter are
thermally activated. This allows to determine a crossover
temperature $T\sim 0.4 \omega_0$ where the polaron crosses over from coherent
to hopping-like motion, which is believed to
hold at strong coupling or in the large phonon  frequency regime.

This crossover can also be observed in the one-electron Green's function.
To this purpose,
we show the spectra in the intermediate coupling regime for increasing $T$
(see fig.  16). The effect is twofold: first,
as is already known from the atomic limit\cite{Mahan},
(see eq. \ref{atomicGreenT}), polaron peaks appear at
negative energies. Secondly, scattering by thermally populated phonon
states causes finite lifetime effects, in addition to the zero-temperature
scattering processes (see Section IV B), as it can be seen by inspecting
the imaginary part of the self-energy.
Indeed, one notices that the negative-$n$ subbands are incoherent.
Moreover, at low $T$, a small peak appears in $Im\Sigma$, close to the upper
boundary of the $n=0$ polaron band. For larger $T$, some spectral weight
develops in this region and contributes to gradually broaden this band,
until the gap eventually disappears. {\it This effect is enhanced in the
vicinity of a
self-energy divergency}, which is extremely sensitive to thermal
disorder. This gives rise to an inhomogeneous
broadening of the main polaron band.

By evaluating the number of coherent states (with $Im \Sigma \ll t$)
within the low energy subbands in the intermediate coupling regime, one can
qualitatively confirm the validity of Holstein's prediction for the crossover
temperature, i.e. $T\sim 0.3-0.4 \omega_0$. On the other hand, the high energy
part of the spectra is slightly smoothened by the temperature.

A different scenario holds
for other values of the
parameters. Results are presented in
figures  17, 18, 19 where the low-energy
part of the electronic spectra is
shown for the same parameter values as in
figs.  10, 11, 12,
for $T = 0.4 \omega_0$.
One notices that in the high phonon frequency
regime, the gaps
between polaron bands exist even at high temperatures, and temperature weakly
affects the overall shape of the positive $n$-th order polaron subbands (see
fig. 17). On the opposite, for low phonon frequencies, the shape of the polaron
subbands is drastically modified (see fig.  19): the spectral weight of
each of them is roughly conserved (for not too large $T$), but they are
noticeably enlarged, and consequently their height diminishes. One must
underline that at those temperatures, the gaps are not destroyed by thermal
fluctuations, apart from vanishingly small spectral density tails.

\section{Conclusion}

We have shown how the dynamical mean-field theory can be successfully used
to solve the single polaron problem at any temperature. The form of the
propagator as a continued fraction expansion, together with the
self-consistency condition for the non-interacting local impurity
propagator, shows that this theory yields an {\it analytic} solution of the
problem, even if some elementary numerics is required to obtain the full
spectra. We have also presented in Section II some results in the limiting
regimes in the infinite dimensional case in order to show that the use of
a semi-elliptic free DOS gives sensible results, in agreement with the
usual three dimensional solutions. This gives confidence that the results
from the dynamical mean-field theory are quite reliable as reflecting the
actual physics in dimension larger than 2. On the contrary, in the
one-dimensional case, one or more gaps are always present in the spectrum, i.e.
polarons form at any coupling \cite{Marsiglio}.

The properties we find for the ground state are in agreement with the
conventional wisdom, in particular the absence of a phase transition,
as soon as quantum fluctuations of the lattice are taken into account.
The crossover gets
sharper as the phonon frequency is decreased for fixed $\lambda$, and
becomes an abrupt transition from a free electron state to a small polaron
state at $\gamma = 0$. This comes directly from the first order nature of
the transition in the adiabatic limit, clearly displayed by the direct
solution.

Beyond the ground state properties, the full electronic spectrum clarifies
the nature of the polaron crossover. For low and intermediate phonon
frequencies, as the coupling is increased, several energy gaps successively
open up in the spectrum, leaving at low energies coherent or quasi-coherent
polaronic subbands. These bands can be followed up to very large couplings
or large phonon frequencies towards the atomic limit. Therefore they are
the manifestation of nonperturbative processes, as is clearly demonstrated
by the interpretation of the CFE in terms of a diagrammatic expansion.
Thus, even if the ground-state properties show a continuous crossover, some
higher energy features show a qualitative change of behaviour, the phenomenon
of gap opening.

Let us make more precise the ''lattice'' interpretation of these polaronic
subbands. Once the self-energy $\Sigma(\omega)$ is known, and for a given
non-interacting lattice DOS, an effective dispersion relation can be found
for each well-separated subband, by finding the poles of the spectral
function eq.(\ref{spectral-function}).  This corresponds to what is
commonly understood as polaron states, i.e. bound states of an electron
with a phonon cloud, a number $n$ of phonons being excited in addition.
For a translationally invariant finite dimensional lattice
the quantum numbers underlying the local (integrated)
spectral function are the {\it total} momentum $K$ of the
polaron\cite{lavorone} and the excited phonon number
$n$. This obviously holds only if the subbands are coherent,
i.e. only for low enough $n$, according to the values
of $\lambda$ and $\gamma$.

The low phonon frequency regime exhibits in the
crossover region a coexistence of extremely narrow
polaronic subbands at low energy and a broad featureless continuum at
high energy. While the low-energy features directly follow from the
atomic or high phonon frequency limit, the continuum is directly related to
the adiabatic solution. The behaviour near the critical
point ($\lambda = \lambda_c, \gamma=0$) can be understood as follows:
increasing $\gamma$, the discontinuity in the low energy properties becomes
a sharp crossover, due to very weak coherent tunneling between small
polaron states and quasi-free states of the equivalent impurity model.
Translated into the language of the lattice problem, it leads to the
emergence of coherent ''heavy'' polaron quasiparticle states
("resonances"). However, the high energy part of the spectrum does not reveal
any qualitative change in this region of parameters.

This behaviour demonstrates how the usual concept of "adiabaticity", as
commonly employed in metals for small $\omega_0$, fails in the present
problem: the sharp transition at $\lambda_c$ and the occurrence of
extremely narrow polaron features indicates on the contrary the occurrence
of an ''adiabaticity catastrophe''. This is due to the relevance of high
order vertex corrections in the perturbation expansion.

In this work the application of the LISA approach has
been limited to the one-particle propagator. In fact, no exact procedure
has been found yet to calculate analytically
for instance two particle propagators.
This would allow to compute
the (dc and optical) conductivity of polarons.
On the other hand, a better understanding
of the polaron crossover could be gained by calculating the dynamical
electron-lattice correlation function.

Let us now comment about extensions of this technique to finite electron
density. As mentioned before, the CFE expansion keeps for the
self-energy the same functional structure as that of the atomic limit.
Thus it is equivalent to a CPA approach which
could be extended at finite densities.
It is however easy to show (by perturbation expansion)
that the CPA approximation fails even
at the first non-vanishing order in the density. As far as spectral
properties are concerned, the CPA failure is
particularly evident in the low energy part of the spectrum,
as is well known from
equivalent approximations for the Hubbard model.
Nevertheless, our main result i.e. the coexistence of low energy
coherent and high energy incoherent
structures is a picture which should be qualitatively
preserved at least at low carriers densities.

Finally, although at zero density it is impossible to access the correlation
functions involved in the calculation of the optical conductivity, let us
mention that our results suggest an interpretation of the infrared
response of oxide  superconductors in the insulating phase. In these
materials\cite{Calvani}, absorption spectra for low carrier densities exhibit a
discrete set of narrow peaks at low energies plus an incoherent background at
higher energies.  From our point of view, this could be ascribed to
multiphonon excitations in the intermediate phonon frequency regime.

\section{Acknowledgements}

S. C. acknowledges financial support from Universit\'e Joseph Fourier,
Grenoble and hospitality of Laboratoire d'Etudes des Propri\'et\'es
Electroniques des Solides. We are grateful to S. Caprara, C. Castellani, A.
Georges, M. Grilli, F. Marsiglio and J. Ranninger for stimulating discussions
or comments

\section{Appendix A: The adiabatic solution}

In this appendix we will solve the problem of one electron moving in an
infinite coordination {\it static} Bethe lattice. To introduce the
adiabatic limit (static lattice) one notices that the free Einstein phonon
Hamiltonian can be written as
\begin{equation}
H_{ph} = \sum_i \frac{P^2_i}{2M}+\frac{1}{2}M \omega^2_0 X_i
\end{equation}
where $P_i$ are the impulses, $X_i$ are the coordinates of the ionic
motion, $M$ is the ionic mass and $\omega_0$ the frequency of
each oscillator.
The adiabatic limit is achieved as $M\rightarrow\infty$ {\it keeping
$k=M\omega^2_0$ constant}.

The electron-phonon interaction can be written as
\begin{equation}
H_{ph} = - g^\prime \sum_i n_i X_i
\end{equation}
where the coupling constant of hamiltonian eq. (\ref{Holstein-model})
is given in terms of $g^\prime$ by $g=g^\prime/\sqrt{2M\omega_0}$.
The polaron energy $\epsilon_p = -g^2/\omega_0 = -g^{'2}/2k$
is then a well defined quantity in the adiabatic limit.

To perform an adiabatic calculation we have to do the following
steps:

i) calculate the electronic energy for a given set of ionic
deformations  $X_i$

ii) minimize the {\it total} energy i.e. electronic energy {\it plus} lattice
elastic energy with respect to the parameters $X_i$.

An essential relation which is useful in the adiabatic limit comes
from the application of the Hellmann-Feynman theorem to the ground
state of the static lattice. By deriving the ground state energy of
the Holstein model with respect to the lattice deformation $X_i$ one
obtains
\begin{equation}
\label{X-n-adiab}
X_i = \frac{g^\prime}{M\omega^2_0} <n_i>
\end{equation}
>From this relation it follows that in the case of a single electron,
due to charge conservation, only two situations are possible:

a) delocalized solution with $X_i=0$ everywhere since the total
charge density per site is zero in the thermodynamic limit.

b) localized solution with some finite $X_i\neq 0$ around one given
site.

Therefore, for a single electron one is restricted to study
electron energies for two different classes of ionic
deformations. Notice that the results quoted above are valid at
any lattice dimensionality.

In the case of the Bethe lattice, a Dyson equation for the local propagator
can be written in the  adiabatic limit
\begin{equation}
\label{eq-motion-adiab}
(\omega+g^\prime X_i -\frac{t^2}{4}\sum_j G_{j,j} ) G_{i,i}  = 1
\end{equation}
then the electronic energy can be derived from the knowledge of the
adiabatic electron propagator which is the solution of eq.
(\ref{eq-motion-adiab}) for a given set of deformations $\{ X_i\}$.

In the case a) (delocalized solution) the solution is trivially the
free electron propagator in the Bethe lattice. The ground state
electron energy is then $E_{el}=-t$.

In the case b) (localized solution) the $d\rightarrow\infty$ limit together
with
eq. (\ref{X-n-adiab}) implies that only one site is appreciably
distorted.  Calling $0$ the center-of-deformation site it is easy to
see that  the nearest neighbor deformation is of the order $1/d$, the
next nearest neighbor deformation is $1/d^2$ and so on, so that
the total charge can be spread on several shells of neighbors even in
the $d\rightarrow\infty$ limit.  The main simplification of the
$d\rightarrow\infty$ limit
is then that the elastic energy {\it is solely determined by the
$0$-site deformation}, for it depends on $X^2_i$. Consequently we
have two kinds of local propagators: one which propagates the
electron from site $0$ back to site $0$, which depends upon the
deformation
\begin{equation}
\label{Goo-ad}
G_{0,0} =  \frac{1}{\omega+g^\prime X_0-\frac{t^2}{4}G}
\end{equation}
and one which propagates the electron from and to any other site,
which is free
\begin{equation}
\label{G11-ad}
G = \frac{1}{\omega-\frac{t^2}{4}G}
\end{equation}
In the case b) a pole in the local Green's function of eq. (\ref{Goo-ad})
emerges out of a band. The position of such a pole determines
the electronic ground state energy. Using eqs. (\ref{Goo-ad},
\ref{G11-ad}) we get the following equation
\begin{equation}
E_{el} + g^\prime X_0 -\frac{1}{2}
Re (E_{el}-\sqrt{E^2_{el}-t^2}) = 0
\end{equation}
which solution is
\begin{eqnarray} E_{el}/t = -1 \;\;\;\;\; \mbox{for} \;\;\;\;\;x<1/2 \\
E_{el}/t = -\frac{1}{4x}-x \;\;\;\;\; \mbox{for}\;\;\;\;\; x>1/2
\end{eqnarray}
where $x=g^\prime X_0/t$. We see that near the zero deformation state
$x=0$ the two possible solutions a) and b) coincide. The
total potential  (rescaled to the hopping energy $t$) we need to
minimize in order to get the ground state energy is obtained by
adding to the solution a) and b) the elastic term (which is zero in
the a) case). Namely
\begin{equation}
\label{E-deloc}
V^{(a)}_{ad}=\frac{x^2}{4\lambda}-1
\end{equation}
while in the b) case
\begin{eqnarray}
\label{E-loc1}
V^{(b)}_{ad}=\frac{x^2}{4\lambda}-1\;\;\;\;\;\mbox{for}\;\;\;\;\;
 x<1/2\\
\label{E-loc2}
V^{(b)}_{ad}=\frac{x^2}{4\lambda}+\frac{1}{4x}+x\;\;\;\;\;\mbox{for}
\;\;\;\;\;x>1/2
\end{eqnarray}
the result is plotted in fig. 3.
There are three different regimes determined by the following critical
values for the coupling constant $\lambda$:
$\lambda^\prime_c=0.650..$ and $\lambda_c=0.844..$

i) $\lambda<\lambda^\prime_c$ The delocalized solution is the stable
minimum.

ii) $\lambda^\prime_c<\lambda<\lambda_c$ The delocalized solution is
the stable minimum  coexisting with a metastable minimum in the
potential $V^{(b)}$, characteristic of the localized solution.

iii) $\lambda>\lambda_c$ The localized solution is the stable
minimum.

In the latter case the the delocalized solution corresponds to
a continuum of unrenormalized excited adiabatic states.  (shaded area
in the lower panel of  fig. 14).

By derivatives of the ground state energy we obtain
the relevant properties of the adiabatic ground state as functions of the
coupling constant.
All these functions can be expressed in terms of the derivative of
the ground state energy $E_0=V_{ad}(X=X_{min})$
with respect to the scaled coupling parameter $x$, namely
$\Delta = dE_0 / d x$.

The local electron-displacement correlation function, the electron kinetic
energy
are determined by deriving the ground state energy (Section II)
\begin{equation}
C_0/2\alpha =-\Delta
\end{equation}
\begin{equation}
E_{kin}/t = -E_0+x\Delta
\end{equation}
Finally, the elastic energy is calculated as a derivative
with respect to $\omega_0$ and scaled to
that parameter so as to get a finite result in the  adiabatic limit
\begin{equation}
E_{ph}/\omega_0 = -x\Delta
\end{equation}

%

\section*{Figure Captions}

\begin{description}

\item[Fig. 1] {A schematic plot of the
regions of parameters' space
($\lambda$,$\gamma$) for the Holstein model in an infinite dimensional Bethe
lattice.
Below the dashed line ($\alpha^2=1$) multiphonon processes are important. On
the bold horizontal axis the adiabatic limit holds. The point in the upper
right corner is reached at the atomic $t=0$ limit. In the shaded area
perturbation theory (small $\lambda$) or Holstein's approximation
(large $\gamma$) are valid. Notice that perturbation theory extends its
validity up to the adiabatic critical value for localization $\lambda_c$
(see text).}

\item[Fig. 2]{The CFE expansion diagrams
obtained by a truncation of the CFE at the first stage
and some of those obtained at the second stage. Diagrams a) and b),c)
represent respectively the second and fourth order perturbation theory
terms. In this case internal propagators are assumed to be free
${\cal G}_0$ .}

\item[Fig. 3]{The adiabatic potential for
a) $\lambda<\lambda^\prime_c$,
b) $\lambda^\prime_c <\lambda<\lambda_c$,
c) $\lambda>\lambda_c$ (see text).}

\item[Fig. 4]{Ground state energy versus
$\lambda$ for three different values of the parameter
$\alpha^2=1,2,5$ (triangles, asterisks, diamonds).  Continuous line
is the adiabatic limit, dashed line the strong coupling result.}

\item[Fig. 5]
{Polaron effective mass in units of the bare electron mass versus
$\lambda$ for three different values of the parameter
$\alpha^2=1,2,5$ (triangles, asterisks, diamonds). Arrows mark the
Holstein's approximation result $\exp (\alpha^2)$}.

\item[Fig. 6]
{Polaron kinetic energy
versus $\lambda$ for three different values of the parameter
$\alpha^2=1,2,5$ (triangles, asterisks, diamonds).
Continuous line is the adiabatic limit result.}

\item[Fig. 7]
{Electron-phonon local correlation function scaled with the strong
coupling result $2\alpha$ versus $\lambda$ for three different values
of the parameter $\alpha^2=1,2,5$ (triangles, asterisks, diamonds).
Continuous line is the adiabatic limit result.}

\item[Fig. 8]
{Average number of phonons in the ground state versus
$\lambda$ for three different values of the parameter
$\alpha^2=1,2,5$ (triangles, asterisks, diamonds). Arrows mark the
strong coupling result  $\alpha^2$.}

\item[Fig. 9]{Both ground-state and spectral properties
are summarized in the $\lambda$,$\gamma$ plane.
a) Effective mass isolines
(from left to right $m^*$=1.1,1.2,1.3,1.5,2,5,20).
In this picture a curve with constant $\alpha^2$ is a straight line
starting at the origin (see also fig.  1).
b) The number of gaps in the electron spectral
density. Near the adiabatic limit all the
lines which separate regions of equal number of gaps
collapse to $\lambda_c$.
On the right of the dashed line
at least one point in which the self-energy diverges appears.}

\item[Fig. 10]{Spectral density (continuous line)
and imaginary part of the self-energy (dashed line) in the large phonon
frequency regime $\gamma=2$, for $\lambda=a) 0.08,b) 0.75,c) 4.0$.
In this and the following spectra the energies
are expressed in units of $t$.}

\item[Fig. 11] {Spectral density (continuous line)
and imaginary part of the self-energy (dashed line) in the intermediate
phonon frequency regime $\gamma=0.5$, for $\lambda=a) 0.4,b) 1.0,c) 2.0$.}

\item[Fig. 12]{Spectral density (continuous line)
and imaginary part of the self-energy (dashed line) in the low phonon
frequency regime $\gamma=0.125$, for $\lambda=a) 0.7,b) 1.0,c) 2.0$.
The delta-like divergencies of the
self-energy near
the first four tiny bands are not reported in this figure
(see fig. 13).}

\item[Fig. 13] {Spectral density and $Im\Sigma$ with the
parameters of fig.  12. The low energy part of the spectrum is
shown, illustrating the $n$-th substructures with $n=1,..,4$ (resp. 0,..,3
excited phonons).  For each
given $n\ge 1$, the coherent and incoherent states are separated by a gap.
The delta peaks in $Im\Sigma$ are revealed
by adding a small imaginary part to the frequency $\omega$.
The width of each structure
has been reported using different scales to compare the shape of each peak.
The bandwidths of the
four coherent peaks are respectively from left to right:
$7.2\times10^{-7}$,
$9.8\times10^{-6}$, $8.33\times10^{-5}$, $5.0134\times10^{-4}$.
The width of the incoherent structures are from left to right
$6.1\times10^{-7}$,
$9.67\times10^{-6}$,
$8.323\times10^{-5}$.}

\item[Fig. 14] {In the upper panel is shown the
spectral density (continuous line) near the adiabatic limit
($\lambda=2$,$\gamma=0.125$) compared with the adiabatic strong coupling
result (dashed line).
In the lower panel continuous line is the ground state
adiabatic total energy as a function of the lattice displacement
(see fig. 3), dashed
line represents the lowest excited adiabatic level which is at the bottom of
a continuum (shaded area). The zero-point energy is omitted here.
See also appendix A.}

\item[Fig. 15] {Evolution of the spectral density and
imaginary part of the self-energy at low energy
for growing $\lambda$ (upper panel)
for $\gamma=0.25$ and, from left to right:
$\lambda=0.7,0.78,0.84,0.9$. The lower panel shows
$\omega-Re\Sigma$.}

\item[Fig. 16] {Evolution of the low energy
spectral density and
imaginary part of self-energy with temperature in the intermediate regime
$\gamma=0.25$ and $\lambda=1.0$.
>From left to right $T=0,0.2,0.4,0.8$. Arrows mark points in which the
self-energy diverges at $T=0$.}

\item[Fig. 17] {Spectral density (continuous line) and imaginary
part of self-energy (long dashed line) for $\gamma=2$ at $T=0.4$.
The spectral density at $T=0$  is shown for comparison (short dashed line).
Panels a) to c) refer to the same $\lambda$'s as fig.  10.}

\item[Fig. 18] {Spectral density (continuous line) and imaginary
part of self-energy (long dashed line) for $\gamma=0.5$ at $T=0.4$.
The spectral density at $T=0$  is shown for comparison (short dashed line).
Panels a) to c) refer to the same $\lambda$'s as fig.  11.}

\item[Fig. 19] {Spectral density (continuous line) and imaginary
part of self-energy (long dashed line) for $\gamma=0.125$ at $T=0.4$.
The spectral density at $T=0$  is shown for comparison (short dashed line).
Panels a) to c) refer to the same $\lambda$'s as fig.  12.}

\end{description}


\begin{references}
\bibitem{adiabatic} For $d>4$
the shape of the band enters in the determination of  the effective
mass which increases with $\lambda$ as $\lambda \gamma$.
\bibitem{adiabatic2} The case $d=2$ is marginal and we cannot gain any
insight from the 2nd order perturbation theory.
\bibitem{alpha2-fix} Notice that if $\alpha^2$ is fixed
increasing $\lambda$ means going toward a large phonon frequency
limit.
\bibitem{strong-adiabatic} Notice that in this limit $\lambda\rightarrow\infty$
while in the adiabatic limit of appendix A $\lambda$ is finite.
\end{references}
\end{document}